\documentclass[a4paper,11pt]{article}
\usepackage{pos}
\usepackage{acronym}
\usepackage{scalefnt}
\usepackage{slashed}
\usepackage{dsfont}
\usepackage[capitalize]{cleveref}
\newcommand{\fig}[1]{Fig.\,\ref{#1}}

\newcommand{\ren}{\text{\abbrev{R}}}

\newcommand{\bare}{\text{\abbrev{B}}}
\newcommand{\dd}{\text{d}}
\newcommand{\dderiv}[2]{\frac{\dd #1}{\dd #2}}
\newcommand{\nf}{n_\text{f}}

\newcommand{\msbar}{\ensuremath{\overline{\text{\abbrev{MS}}}}}

\newcommand{\order}[1]{\ensuremath{{\cal O}(#1)}}
\newcommand{\lmut}{l_{\mu t}}

\newcommand{\EulerGamma}{\gamma_\text{E}}
\newcommand{\ep}{\epsilon}

\newcommand{\calo}{\mathcal{O}}
\newcommand{\tcalo}{\tilde{\mathcal{O}}}

\newcommand{\abbrev}[1]{{\scalefont{.9}#1}}
\newcommand{\citere}[1]{Ref.\,\cite{#1}}
\newcommand{\citeres}[1]{Refs.\,\cite{#1}}
\newcommand{\myacrodef}[3]{\acrodef{#2}{#3}\newcommand{#1}{\ac{#2}}}
\myacrodef{\sftx}{SFTX}{small-flow-time expansion}
\myacrodef{\gff}{GFF}{gradient-flow formalism}
\myacrodef{\qcd}{QCD}{Quantum Chromodynamics}
\myacrodef{\lo}{LO}{leading order}
\myacrodef{\nlo}{NLO}{next-to-leading order}
\myacrodef{\nnlo}{NNLO}{next-to-next-to-leading order}

\title{The gradient flow at higher orders in perturbation theory}

\author{Robert Harlander}

\affiliation{TTK, RWTH Aachen University\\
  Sommerfeldstr.~16, Aachen, Germany}

\emailAdd{harlander@physik.rwth-aachen.de}

\abstract{Various results for higher-order perturbative calculations in
  the gradient-flow formalism are reviewed, including the gradient-flow
  beta function and the small-flow-time expansion of the hadronic vacuum
  polarization and the energy-momentum tensor. In addition, the
  \textit{strategy of regions} is outlined in order to obtain systematic
  expansions of gradient-flow integrals, for example at large and small
  flow times.}

\FullConference{%
 The 38th International Symposium on Lattice Field Theory, LATTICE2021
  26th-30th July, 2021
  Zoom/Gather@Massachusetts Institute of Technology
  --- preprint TTK-21-49
}

\begin{document}
\maketitle


\section{Introduction}

\qcd\ is a theory with many different facets. So far, quantitative
phenomenological results have been obtained mostly in either the
strong-coupling regime using lattice regularizations or in the
weak-coupling regime where perturbation theory is applicable. Both
calculational approaches are highly evolved in
themselves. Cross-fertilization is often hindered by the inherently
different treatment of ultraviolet divergences in these two
calculational approaches.

The \gff\ may provide an excellent opportunity to change this
situation. It represents a \abbrev{UV} regularization scheme which can
be implemented both on the lattice and in perturbation theory. This
contribution reviews a number of concrete examples where such a
cross-fertilization could be achieved, and where the perturbative
calculations have been performed beyond next-to-leading order in
perturbation theory. Furthermore, the application of the
\texttt{strategy of regions} to gradient-flow integrals is presented,
which allows to obtain systematic expansions in dimensionless
parameters.


\section{The perturbative gradient flow}

In the \gff, one defines flowed fields $B^a_\mu=B^a_\mu(t)$ and
$\chi=\chi(t)$ as solutions of the
equations\,\cite{Luscher:2010iy,Luscher:2013cpa} (see also
\cite{Narayanan:2006rf,Luscher:2009eq})
\begin{equation}
  \begin{split}
    \partial_t B^a_\mu &= \mathcal{D}^{ab}_\nu G^b_{\nu\mu} + \kappa
    \mathcal{D}^{ab}_\mu \partial_\nu B^b_\nu\,,\\ \partial_t \chi
    &= \Delta \chi - \kappa \partial_\mu B^a_\mu T^a \chi\,,\\ \partial_t
    \bar \chi &= \bar \chi \overleftarrow \Delta + \kappa \bar
    \chi \partial_\mu B^a_\mu T^a\,.
    \label{eq:flow}
  \end{split}
\end{equation}
The initial conditions supplementing these differential equations
establish the contact to regular \qcd:
\begin{equation}
  \begin{split}
    B^a_\mu (t=0) = A^a_\mu\,,\qquad \chi (t=0)= \psi\,,
    \label{eq:bound}
  \end{split}
\end{equation}
where $A^a_\mu$ and $\psi$ are the regular gluon and quark fields,
respectively, and
\begin{equation}\label{eq:dleftright}
  \begin{split}
    \mathcal{D}^{ab}_\mu &= \delta^{ab}\partial_\mu - f^{abc}
    B_\mu^c\,,\qquad \Delta = (\partial_\mu + B^a_\mu T^a) (\partial_\mu
    + B^b_\mu T^b)\,,\\ G_{\mu\nu}^a &= \partial_\mu B_\nu^a -
    \partial_\nu B_\mu^a + f^{abc}B_\mu^bB_\nu^c\,.
  \end{split}
\end{equation}
The arbitrary parameter $\kappa$ will be set equal to one in the
following.

Our practical implementation of the \gff\ in perturbation theory follows
the strategy developed in \citere{Luscher:2011bx}. It leads to Feynman
rules resembling those of regular \qcd, but supplemented by flow-time
dependent exponentials in the propagators. In addition, the
flow equations are reflected through so-called flow lines which couple
to the flowed quarks and gluons via flowed vertices. The latter involve
integrations over finite intervals of the flow-time variables.

A systematic method how to handle the corresponding Feynman diagrams and
integrals through three-loop level has been introduced in
\citere{Artz:2019bpr}. It is based on
\texttt{qgraf}\,\cite{Nogueira:2006pq} for the generation of Feynman
diagrams,
\texttt{FORM}\,\cite{vanRitbergen:1998pn,Vermaseren:2000nd,Kuipers:2012rf}
for the algebraic manipulation of the resulting amplitudes,
\texttt{Kira+FireFly}\,\cite{Maierhofer:2017gsa,Maierhofer:2018gpa,
  Klappert:2019emp,Klappert:2020aqs,Klappert:2020nbg} for the reduction
of the Feynman integrals to master integrals, and
\texttt{q2e/exp}\,\cite{Harlander:1998cmq} for interfacing all of these
programs. The master integrals can be calculated by following the method
outline in \citere{Harlander:2016vzb}, for example.


\section{Gluon condensate, quark condensates, and gradient-flow beta function}


\subsection{Gluon condensate and gradient-flow beta function}

\begin{figure}
  \begin{center}
    \begin{tabular}{cc}
      \mbox{\includegraphics[%
        viewport=20 10 300 220,
        width=.45\textwidth]%
                          {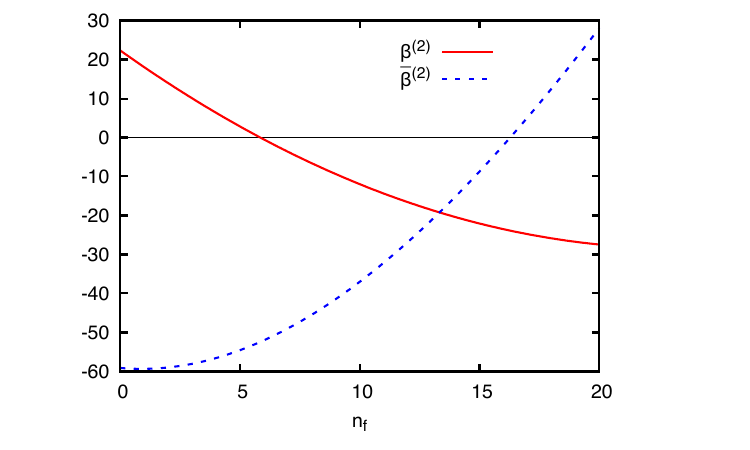}}
                          &
          \mbox{\includegraphics[%
            viewport=20 10 300 220,
            width=.45\textwidth]%
            {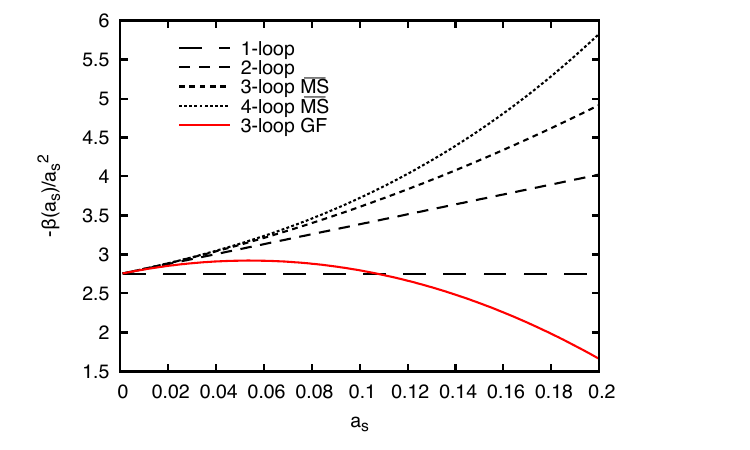}}\\
          (a) & (b)
    \end{tabular}
    \parbox{.9\textwidth}{
      \caption[]{\label{fig:beta}\sloppy (a) Three-loop coefficient of
        the $\beta$ function in the \msbar\ (solid-red) and the
        gradient-flow scheme (dashed-blue). (b) $\beta$ function in the
        \msbar\ scheme and the gradient-flow scheme for $n_f=0$
        ($\mathrm{a}_s\equiv \alpha_s/\pi$). }}
  \end{center}
\end{figure}

The first quantity considered at finite flow time was the gluon
condensate in massless \qcd\,\cite{Luscher:2010iy}. Its perturbative
expansion reads
\begin{equation}\label{eq:iron}
  \begin{split}
    &\langle G_{\mu\nu}^a(t)G_{\mu\nu}^a(t) \rangle = 3
    \frac{\alpha_s(\mu)}{\pi t^2}\bigg[ 1 +
      \frac{\alpha_s(\mu)}{4\pi}\left(e_{10} + 4\beta_0l_{\mu t}\right)
      \\& + \left(\frac{\alpha_s(\mu)}{4\pi}\right)^2 \left(e_{20} +
      8\left(e_{10}\,\beta_0 + 2\,\beta_1\right)l_{\mu t}+
      16\,\beta_0^2\,l_{\mu t}^2\right)\bigg] + \cdots\quad \equiv\quad
    3\frac{\bar\alpha_s(t)}{\pi t^2}\,,
  \end{split}
\end{equation}
where $\lmut \equiv \ln 2\mu^2 t+\EulerGamma$ with Euler's constant
$\EulerGamma= 0.577\ldots$, and $\alpha_s(\mu)$ is the strong coupling
in the \msbar\ scheme which obeys
\begin{equation}\label{eq:hare}
  \begin{split}
    \mu^2\dderiv{}{\mu^2}\frac{\alpha_s(\mu)}{\pi} = \beta(\alpha_s(\mu))\,,
\quad\text{with}\quad
    \beta(\alpha_s) =
    -\sum_{n=0}^\infty \left(\frac{\alpha_s}{\pi}\right)^{2+n}\beta_n\,.
  \end{split}
\end{equation}
In \qcd, the first three coefficients of the $\beta$ function in the
\msbar\ scheme read
\begin{equation}\label{eq:jaws}
  \begin{split}
    \beta_0 &= \frac{11}{4}-\frac{n_f}{6}\,,\qquad
    \beta_1 = \frac{51}{8}-\frac{19}{24}\,n_f\,,\\
    \beta_2 &= \frac{2857}{128} - \frac{5033}{1152}\,n_f +
    \frac{325}{3456}\,n_f^2
    \approx 22.3 - 4.37\,n_f + 0.094\,n_f^2\,.
  \end{split}
\end{equation}
Setting $t$ to some multiple of $\mu^2$, i.e.\ $t = \mu^2/\rho$, acting
with $\mu^2 \dd/\dd\mu^2$ on \cref{eq:iron}, and iteratively replacing
$\alpha_s$ by $\bar\alpha_s$ according to \cref{eq:iron}, one finds the
evolution of the gradient-flow coupling:
\begin{equation}\label{eq:glob}
  \begin{split}
    \mu^2\dderiv{}{\mu^2}\frac{\bar\alpha_s(\mu)}{\pi} =
    \bar\beta(\bar\alpha_s(\mu))\,,
\quad\text{with}\quad
    \bar\beta(\bar\alpha_s) =
    -\sum_{n=0}^\infty \left(\frac{\bar\alpha_s}{\pi}\right)^{2+n}\bar\beta_n\,.
  \end{split}
\end{equation}
The first two coefficients are universal, i.e.\ $\bar\beta_0=\beta_0$
and $\bar\beta_1=\beta_1$, while
\begin{equation}\label{eq:horn1}
  \begin{split}
    \bar\beta_2 =
    \beta_2-\frac{1}{4}e_{10}\beta_1+\frac{1}{16}\left(e_{20}-e_{10}^2\right)
    \beta_0
     = -59.1 - 0.536\,n_f + 0.304\,n_f^2 - 0.0030\,n_f^3\,.
  \end{split}
\end{equation}
Note that the $\rho$-dependence drops out of the $\bar\beta$ function.
The difference to the $\msbar$ value of \cref{eq:jaws} is remarkable.  It
is illustrated in dependence of $n_f$ in \cref{fig:beta}\,(a). The
impact on the \qcd\ $\beta$ function is shown in
\cref{fig:beta}\,(b). One immediately notices that the perturbative
convergence is significantly worse in the gradient-flow scheme than in
the \msbar\ scheme (see also
\citeres{DallaBrida:2016dai,Fodor:2017die,DallaBrida:2019wur,
  Peterson:2021lvb}). It would be interesting to understand the source
of this behavior in order to allow for a precise independent lattice
determination of $\alpha_s(M_Z)$ through the \gff.


%
%


%
\begin{figure}
  \begin{center}
    \begin{tabular}{cc}
          \includegraphics[%
            width=.2\textwidth]%
                          {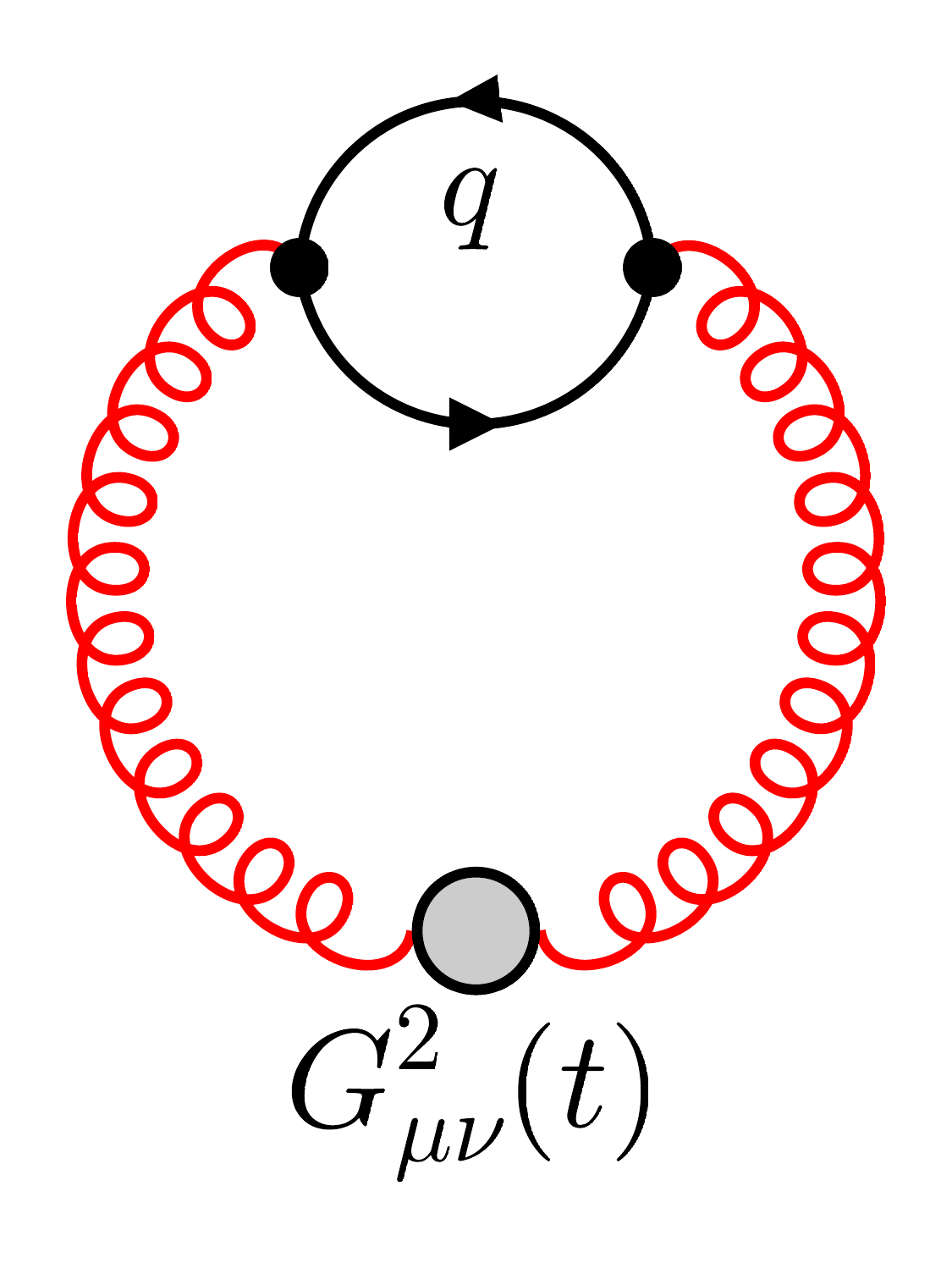}
                          &
          \includegraphics[%
            width=.2\textwidth]%
                          {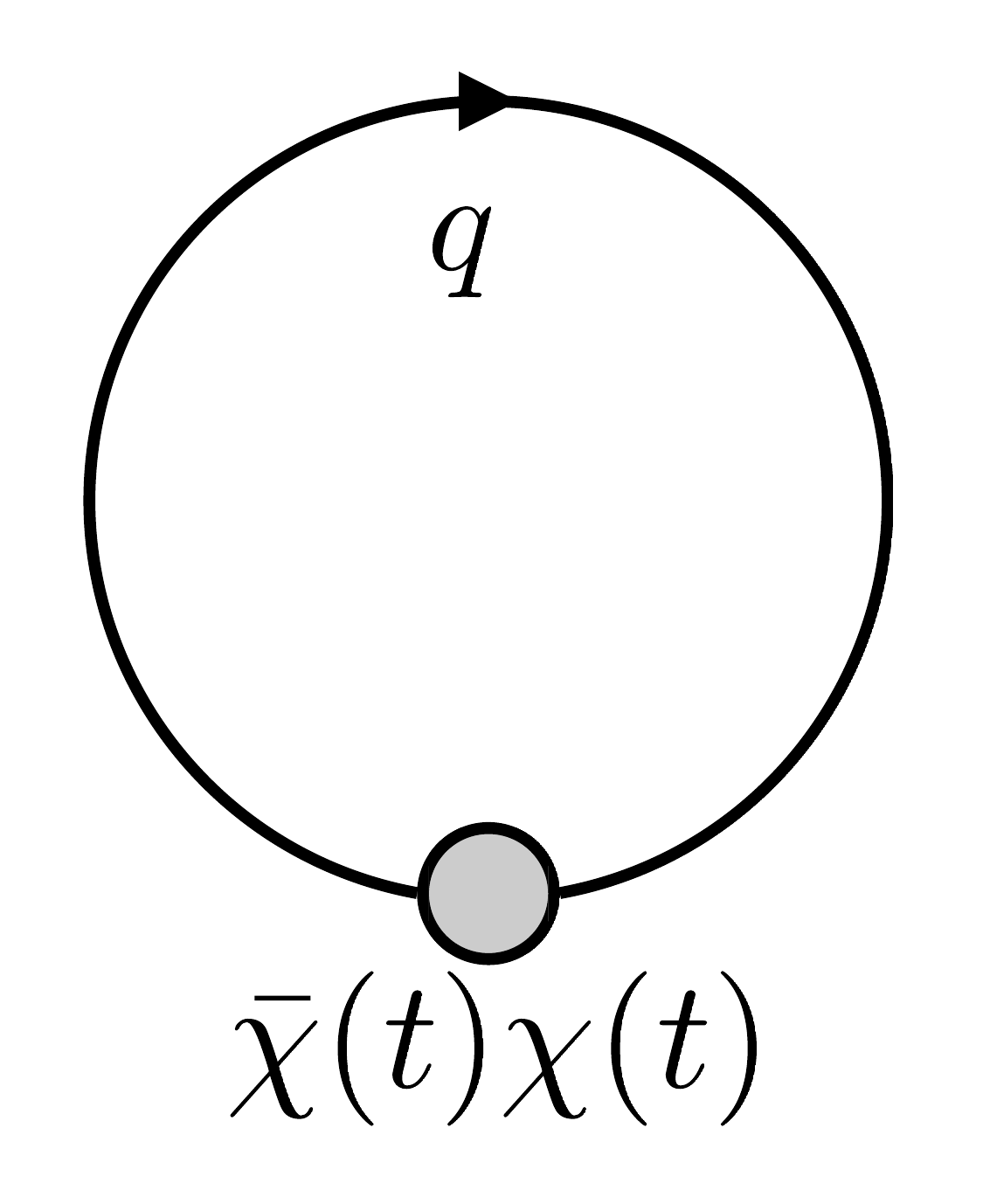}\\
                          (a) & (b)
    \end{tabular}
    \parbox{.9\textwidth}{
      \caption[]{\label{fig:vevs}\sloppy Leading-order contributions for
        the quark mass effects to the gluon and the quark
        condensate. Diagrams produced with
        \texttt{FeynGame}\,\cite{Harlander:2020cyh}.  }}
  \end{center}
\end{figure}
%



\subsection{Quark mass effects}

So far, we have considered the case of massless quarks. For the gluon
condensate, quark mass effects occur only at \nlo\ through the single
Feynman diagram shown in \fig{fig:vevs}\,(a)\,.  They can be taken into
account quite easily by using the well-known one-loop expression for the
two-point function with external gluons. The result has been expressed
in terms of a one-dimensional integral\,\cite{Harlander:2016vzb}. At
higher orders in perturbation theory, approximate results of the mass
effects could be obtained using the so-called \textit{strategy of
  regions}\,\cite{Beneke:1997zp}. To illustrate its application in the
\gff, let us consider the simpler case of the quark condensate, where
mass effects occur already at \lo, see \fig{fig:vevs}\,(b). The exact
mass dependence leads to an incomplete $\Gamma$ function in this
case\,\cite{Luscher:2013cpa}:
\begin{equation}\label{eq:horn2}
  \begin{split}
    S(t) \equiv \langle\bar\chi(t)\chi(t)\rangle = -\frac{3m}{8\pi^2
      t}f(m^2,2\,t)\,,
  \end{split}
\end{equation}
with
\begin{equation}\label{eq:flew}
  \begin{split}
    f(m^2,t)&\equiv (4\pi t)^{D/2} t^{-1}\int_k\frac{e^{-tk^2}}{k^2+m^2}
    =1-m^2t e^{m^2t}\Gamma(0,m^2t)\,,\\
    \int_k&\equiv \int\frac{\dd^D k}{(2\pi)^D}\,,\qquad
    \Gamma(s,x) = \int_x^\infty\dd u\,u^{s-1}e^{-u}\,,
  \end{split}
\end{equation}
where we have dropped terms that vanish as $\ep\to 0$.
Assume that we would like to solve the momentum integral in
\cref{eq:flew} as an expansion around $m^2 \ll 1/t$. Obviously, simply
interchanging the expansion with the integration, which corresponds to
assuming $k^2\gg m^2$ in the integrand, leads to \abbrev{IR}-divergent
integrals:
\begin{equation}\label{eq:cube}
  \begin{split}
    f^{(i)}(m^2,t) &= (4\pi t)^{D/2} t^{-1}\sum_{n=1}^\infty (-m^2)^{n-1}\int_k
    \frac{e^{-tk^2}}{k^{2n}}= \sum_{n=1}^\infty
    (-m^2t)^{n-1}\frac{\Gamma(D/2-n)}{\Gamma(D/2)}= \\& = 1 +
    m^2t\left(\frac{1}{\ep}+1\right)e^{m^2t} - (m^2t)^2 -
    \frac{3}{4}(m^2t)^2 + \ldots\,.
  \end{split}
\end{equation}
On the other hand, considering the region $k^2\sim m^2$, it follows that
$tk^2\ll 1$, so we can expand the exponential:
\begin{equation}
  \begin{split}
    f^{(ii)}(m^2,t)
    &=(4\pi t)^{D/2}t^{-1}\sum_{n=0}^\infty \frac{(-t)^n}{n!}
    \int_k\frac{k^{2n}}{k^2+m^2}
    = m^2t\left[-\frac{1}{\ep}-1 +\EulerGamma+ \ln m^2t
      \right]e^{m^2t}\,.
  \end{split}
\end{equation}
We recall that, despite the fact that the expansion of the integrand is
justified only in the respective region, the momentum integral can be
taken over all values of $k$, because all complementary regions will
combine to scale-less integrals which are discarded in dimensional
regularization.  Combining the two regions, the $1/\ep$ pole cancels and
one finds
\begin{equation}
  \begin{split}
    f(m^2,t)&\stackrel{m^2\ll 1/t}{\rightsquigarrow}f^{(i)}(m^2,t) +
    f^{(ii)}(m^2,t)=\\
    &= 1 + m^2 t\,
    e^{m^2t}\,(\ln m^2t + \EulerGamma) -(m^2t)^2 - \frac{3}{4}(m^2t)^3
    + \order{(m^2t)^4}\,,
  \end{split}
\end{equation}
which agrees with the asymptotic expansion of the explicit expression
given in \cref{eq:flew}.

Now let us consider the opposite case: $m^2\gg 1/t$. We have again two
regions, the first one leading to
\begin{equation}
  \begin{split}
    f(m^2,t)&\stackrel{m^2\gg k^2}\rightsquigarrow \hat{f}^{(i)}(m^2,t)
    = (4\pi t)^{D/2}t^{-1}
    \sum_{n=0}^\infty\frac{(-1)^n}{m^{2(n+1)}}\int_k\,k^{2n}\,e^{-k^2t}=\\
     &=\sum_{n=1}^\infty (-1)^{n+1}n!\, (m^2t)^{-n}\,.
  \end{split}
\end{equation}
The second region is given again by $k^2\sim m^2$, which means that
$k^2t\gg 1$. Its contribution vanishes, because the Taylor series of
$e^{-1/x}$ around $x=0$ is identical to zero. Therefore,
\begin{equation}\label{eq:jodi}
  \begin{split}
    f(m^2,t)\stackrel{m^2\gg 1/t}{\rightsquigarrow} =
    \hat{f}^{(i)}(m^2,t) = \frac{1}{m^2 t} - \frac{2}{(m^2 t)^2} +
    \frac{6}{(m^2 t)^3} + \ldots
  \end{split}
\end{equation}
which again agrees with the Taylor series of \cref{eq:flew} around
$1/(m^2t)=0$.

Of course, our presentation here is only a sketch of the general
idea. At higher orders, one needs to take into account integrations over
flow-time parameters. In the small-$t$ limit, all flow-time integration
variables are bound to be small as well, and the extension to higher
loop order is straightforward. In the large-$t$ limit, however,
integration over flow-time parameters extends over ``large'' and
``small'' regions, and the expansion of the integrand becomes
non-trivial. A general treatment of the strategy of regions for
flow-time integrals at higher orders is thus ongoing work and will be
presented elsewhere.


\section{Hadronic vacuum polarization}

Consider the operator product expansion of the correlator of two vector
currents $j_\mu(x)$ in $n_f$-flavor \qcd\ with a single massive quark
flavor~\cite{Dominguez:2014vca}:
\begin{equation}
  \begin{split}
    \Pi_{\mu\nu}(Q) &\equiv \int\dd^4 x\, e^{iQx} \langle
    Tj_\mu(x)j_\nu(0)\rangle = (-\delta_{\mu\nu} + Q_\mu Q_\nu/Q^2)\Pi(Q^2)
    \\&
    \Pi(Q^2)\stackrel{Q^2\to \infty}{\sim}
    C^{(0),\bare}(Q) + m_\bare^2C^{(2),\bare}(Q) + \sum_{n}
    C^\bare_{n}(Q)\langle\calo_{n}(x=0)\rangle\,,
    \label{eq:ccope}
  \end{split}
\end{equation}
This form reflects the fact that, up to mass dimension two, only the
trivial operators $\mathds{1}$ and $m_\bare^2\mathds{1}$ contribute,
where $m_\bare$ is the bare quark mass.  At mass dimension four, one has
the following set of physical operators (the space-time argument is
suppressed in most of what follows):
\begin{equation}
  \begin{split}
    \calo_{1} &= \frac{1}{g_\bare^2}F^a_{\mu\nu}
    F^a_{\mu\nu}\,,\qquad
    \calo_{2} =
    \sum_{q=1}^{\nf}\bar\psi_q\overleftrightarrow{\slashed{D}}\psi_q\,,\qquad
     \calo_{3} =   m_\bare^4\mathds{1}\,,
    \label{eq:calo}
  \end{split}
\end{equation}
\begin{equation}\label{eq:overF}
  \begin{split}
\mbox{where}\qquad     F^a_{\mu\nu} &= \partial_\mu A^a_\nu - \partial_\nu A^a_\mu +
     f^{abc}A_\mu^bA_\nu^c\,, \qquad\overleftrightarrow{D}\!_\mu =
     \partial_\mu - \overleftarrow{\partial}_\mu + 2A^a_\mu T^a\,.
  \end{split}
\end{equation}
After renormalization of $m_\bare$ and the bare coupling $g_\bare$, the
operator matrix elements on the r.h.s.\ of \cref{eq:ccope} are still
divergent. The divergences can be absorbed into the bare coefficient
functions with the help of operator renormalization:
\begin{equation}\label{eq:ribfo}
  \begin{split}
    \sum_n C_n^\bare\calo_n \equiv \sum_n C_nZ_{nm}\calo_m \equiv \sum_n
    C_n\calo^\ren_m\,.
  \end{split}
\end{equation}
The fact that the classical mass dimension of the operators is the same
as that of the Lagrangian allows one to express the renormalization
matrix $Z$ in terms of the \qcd\ $\beta$ function, the quark mass
anomalous dimension, and the anomalous dimension of the vacuum energy to
all orders~\cite{Spiridonov:1988md}.

The operator product expansion of \cref{eq:ccope} represents a
factorization into long- and short-distance effects. The former are
contained in the matrix elements $\langle\calo_n\rangle$ and their
evaluation requires non-perturbative methods such as lattice \qcd. The
latter are in the coefficient function whose perturbative expressions
are known through \nnlo\ and beyond (see
\citeres{Chetyrkin:1985kn,Chetyrkin:1997qi,Harlander:1997xa}, for
example).  A precise prediction of $\Pi_{\mu\nu}(Q)$ requires full
control of the matching between the two, which is notoriously difficult
due to the different regularization and renormalization schemes.

The \sftx\ provides a potential solution to this problem by unifying the
renormalization scheme for both the coefficient functions and the
operators\,\cite{Luscher:2011bx}. The spectrum of possible applications
is enormous (see
\citeres{Suzuki:2013gza,Makino:2014taa,Hieda:2016lly,Iritani:2018idk,
  Harlander:2020duo,Suzuki:2020zue,Rizik:2020naq,Mereghetti:2021nkt},
for example). The idea is to define flowed operators $\tcalo_n(t)$ by
replacing the regular by flowed fields in \cref{eq:calo}, and then
expressing them in terms of regular operators in the limit $t\to 0$:
\begin{equation}\label{eq:acme}
  \begin{split}
    \tcalo_n(t)&\asymp
    \zeta_n^{(0),\bare}(t)+
    \zeta_n^{(2),\bare}(t)m_\bare^2 +
    \sum_{m}\zeta^\bare_{nm}(t)\calo_m =
    \zeta_n^{(0)}(t)+
    \zeta_n^{(2)}(t)m^2 +
    \sum_{m,k}\zeta_{nk}(t)\calo^\ren_k\,.
  \end{split}
\end{equation}
Here, the symbol $\asymp$ denotes that the relation holds only
asymptotically for $t\to 0$. The coefficients $\zeta_n^{(0)}(t)$,
$\zeta_n^{(2)}(t)$, and $\zeta_{nm}(t)$ are \abbrev{UV} finite. They
have been calculated in \citere{Harlander:2020duo} through \nnlo\ \qcd.
While the $\zeta_{nm}(t)$ depend only logarithmically on $t$,
$\zeta_n^{(0)}(t)$ and $\zeta_n^{(2)}(t)$ behave as $1/t^2$ and $1/t$ as
$t\to 0$. In fact, they simply correspond to the first two terms in the
Taylor expansion of the vevs around $m=0$:
\begin{equation}\label{eq:bebe}
  \begin{split}
    \zeta_n^{(0)}(t) + 
    \zeta_n^{(2)}(t)m^2 =
    \langle\mathcal{O}_n(t)\rangle\Big|_{m=0} + 
    m^2\frac{d}{d m^2}\langle\mathcal{O}_n(t)\rangle\Big|_{m=0}\,.
  \end{split}
\end{equation}
Inverting \cref{eq:acme} and inserting it into \cref{eq:ccope} leads to
\begin{equation}\label{eq:flab}
  \begin{split}
    \Pi(Q) \stackrel{Q^2\to\infty}{\sim}
    C^{(0),\bare}(Q) + m^2C^{(2)}(Q)+
    \sum_{n}\tilde{C}_n(t)\bar{\mathcal{O}}_n(t)\,,
  \end{split}
\end{equation}
\begin{equation}\label{eq:inyl}
  \begin{split}
\mbox{with}\qquad    \tilde{C}_n(t) = \sum_{m}C_m\zeta^{-1}(t)_{mn}\,,\qquad
    \bar{\mathcal{O}}_n(t) = \tcalo_n(t)- \zeta_n^{(0)}(t) -
    m^2\zeta_n^{(2)}(t)\,.
  \end{split}
\end{equation}
Note that power divergences in the limit $t\to 0$ cancel in the
combination $\bar{O}_n(t)$. A precise lattice determination of the
$\langle\bar{\mathcal{O}}(t)$ could thus open the way towards a novel
calculation of the vacuum polarization, and thus independent input for
the lattice determination of hadronic contributions to low-energy
observables such as the muon anomalous magnetic moment.


\section{Energy-momentum tensor}

Dropping terms that vanish either under a \abbrev{BRST} transformation
or by equations of motion, the energy-momentum tensor of \qcd\ takes the
form of an operator product expansion similar to \cref{eq:ccope}:
\begin{equation}\label{eq:tmunu}
    \begin{split}
      T_{\mu\nu} = \sum_{n=1}^4 C_n\mathcal{O}_{n,\mu\nu}\,,
    \end{split}
\end{equation}
\begin{equation}\label{eq:gale}
  \begin{split}
    \mbox{where}\qquad
    \mathcal{O}_{1,\mu\nu} &= \frac{1}{g_\bare^2}G_{\mu\rho}G_{\rho\nu}\,,\qquad
    \mathcal{O}_{2,\mu\nu} =
    \frac{\delta_{\mu\nu}}{g_\bare^2}G_{\rho\sigma}G_{\rho\sigma}\,,\\
    \mathcal{O}_3 &= \bar\psi\left(\gamma_\mu\overleftrightarrow{D}_\nu
    +\gamma_\nu\overleftrightarrow{D}_\mu\right)\psi\,,\qquad
    \mathcal{O}_4 = \delta_{\mu\nu}m_\bare\bar\psi\psi\,.
  \end{split}
\end{equation}
However, as opposed to \cref{eq:ccope}, the ``Wilson coefficients'' $C_n$
in this case are given by simple numerical constants to all orders in
perturbation theory:
\begin{equation}\label{eq:eely}
  \begin{split}
    C_1 = 1\,,\qquad C_2=-\frac{1}{4}\,,\qquad
    C_3 = \frac{1}{4}\,,\qquad C_4=0\,.
  \end{split}
\end{equation}
Furthermore, due to the Ward-Takahashi identities among the $Z_{nm}$,
the energy-momentum tensor is \textit{finite}, in the sense that the
coefficients $C_n$ are not renormalized, i.e.\ $C_n=C_n^\bare$.

Using the \sftx, we write the operators as
\begin{equation}\label{eq:cluj}
  \begin{split}
   \mathcal{O}_n \asymp
   \sum_{m=1}^4\zeta^{\bare,-1}(t)_{nm}\tilde{\mathcal{O}}_m(t)
  \end{split}
\end{equation}
and insert this into \cref{eq:tmunu} to
obtain\,\cite{Suzuki:2013gza,Makino:2014taa}
\begin{equation}\label{eq:cord}
  \begin{split}
    T_{\mu\nu} \asymp \sum_{n=1}^4
    c_n(t)\tilde{\mathcal{O}}_{n,\mu\nu}(t)\,,\qquad
    c_n(t) = \sum_{m=1}^4 C_m\zeta^{\bare,-1}_{mn}(t) = \sum_{m=1}^4
    C_m\zeta^{-1}_{mn}(t)\,.
  \end{split}
\end{equation}
The coefficients $c_n(t)$ are finite even without operator
renormalization. They have been evaluated through
\nnlo~\cite{Suzuki:2013gza,Makino:2014taa,Harlander:2018zpi} and used to
study thermodynamics of
\qcd\,\cite{Iritani:2018idk,Kitazawa:2019otp,Shirogane:2020muc,
  Yanagihara:2020tvs,Taniguchi:2020mgg}.


\paragraph{Acknowledgments.}
I would like to thank Aiman El Asad, Janosch Borgulat, Benedikt Gurdon,
Stefano Palmisano, Fabian Lange, Tobias Neumann, Paul Mork, Antonio
Rago, Joshua Schophoven, and Andrea Shindler for useful and inspiring
conversations, as well as Hiromasa Takaura for pointing out a typo in
\cref{eq:horn2} of the original version of this paper.

\bibliography{harlander}

\end{document}